%% file: main.tex
\documentclass[]{ieeeaccess}

\usepackage{amsmath,amsfonts}
\usepackage{algorithmic}
\usepackage{array}
\usepackage[caption=false,font=normalsize,labelfont=sf,textfont=sf]{subfig}
\usepackage{textcomp}
\usepackage{stfloats}
\usepackage{url}
\usepackage{verbatim}
\usepackage{graphicx}
\usepackage{balance}
\usepackage{pbox}

\usepackage{tabularx}
\usepackage{multirow}
\usepackage{pifont}
\usepackage{graphicx}
\usepackage{amsfonts}
\usepackage{hyperref}
\hypersetup{
   colorlinks=true,
   linkcolor=black,
   anchorcolor=black,
   citecolor=black,
   filecolor=black,
   urlcolor=black
}
\usepackage{arydshln}

\newcommand{\cmark}{{\ding{51}}}
\newcommand{\xmark}{{\ding{55}}}

\usepackage{soul}
\newcommand*{\HL}[1]{#1}

\usepackage{bm}
\makeatletter
\AtBeginDocument{\DeclareMathVersion{bold}
\SetSymbolFont{operators}{bold}{T1}{times}{b}{n}
\SetSymbolFont{NewLetters}{bold}{T1}{times}{b}{it}
\SetMathAlphabet{\mathrm}{bold}{T1}{times}{b}{n}
\SetMathAlphabet{\mathit}{bold}{T1}{times}{b}{it}
\SetMathAlphabet{\mathbf}{bold}{T1}{times}{b}{n}
\SetMathAlphabet{\mathtt}{bold}{OT1}{pcr}{b}{n}
\SetSymbolFont{symbols}{bold}{OMS}{cmsy}{b}{n}
\renewcommand\boldmath{\@nomath\boldmath\mathversion{bold}}}
\makeatother

\def\BibTeX{{\rm B\kern-.05em{\sc i\kern-.025em b}\kern-.08em
    T\kern-.1667em\lower.7ex\hbox{E}\kern-.125emX}}

\begin{document}

\history{Date of publication xxxx 00, 0000, date of current version xxxx 00, 0000.}
\doi{10.1109/ACCESS.2024.0429000}

\title{Domain Adapting Deep Reinforcement Learning for Real-world Speech Emotion Recognition}
\author{
    \uppercase{Thejan~Rajapakshe}\authorrefmark{1}, \IEEEmembership{Student Member, IEEE},
    \uppercase{Rajib~Rana}\authorrefmark{1}, \IEEEmembership{Member, IEEE},
    \uppercase{Sara~Khalifa}\authorrefmark{2}, \IEEEmembership{Member, IEEE},
    \uppercase{Bj\"{o}rn W.\ Schuller}\authorrefmark{3,4}, \IEEEmembership{Fellow, IEEE}
}

\address[1]{School of Mathematics, Physics and Computing, University of Southern Queensland, Toowoomba, QLD 4350, Australia}
\address[2]{School of Information Systems, Queensland University of Technology, Brisbane, QLD 4000, Australia}
\address[3]{Embedded Intelligence for Health Care \& Wellbeing, University of Augsburg, 86159 Augsburg, Germany}
\address[4]{GLAM -- Group on Language, Audio, \& Music, Imperial College London, London SW7 2AZ London, U.K}

\markboth
{T. Rajapakshe \headeretal: Domain Adapting Deep Reinforcement Learning for Real-world Speech Emotion Recognition}
{T. Rajapakshe \headeretal: Domain Adapting Deep Reinforcement Learning for Real-world Speech Emotion Recognition}

\corresp{Corresponding author: Thejan Rajapakshe (e-mail: thejan.rajapakshe@unisq.edu.au).}

\begin{abstract}
Speech-emotion recognition (SER) enables computers to engage with people in an emotionally intelligent way. The inability to adapt an existing model to a new domain is one of the significant limitations of SER methods. To overcome this challenge, domain adaptation techniques have been developed to transfer the knowledge learnt by a model across domains. Although existing domain adaptation techniques have improved the performance of SER models across domains, there is a need to improve their ability to adapt to real-world situations where models can self-tune while deployed. This paper presents a deep reinforcement learning-based strategy (RL-DA) for adapting a pre-trained SER model to a real-world setting by interacting with the environment and collecting continuous feedback. The proposed RL-DA technique is evaluated on SER tasks, including cross-corpus and cross-language domain adaptation scenarios. Our evaluation results show that RL-DA achieves significant improvements of 11\% and 14\% in testing accuracy over a fully supervised baseline for cross-corpus and cross-language scenarios, respectively, in the real-world setting. This technique also outperforms the baseline model's performance for both speaker independent and speaker dependent SER tasks. 
\end{abstract}


\titlepgskip=-21pt

\maketitle

\IEEEdisplaynontitleabstractindextext

%

\input{10_introduction}

\input{20_literature_review}

\input{30_methodology}

\input{35_experimental_setup}

\input{40_results}

\input{45_discussion}

\input{60_conclusion}

\ifCLASSOPTIONcaptionsoff
  \newpage
\fi

\bibliographystyle{IEEEtran}
\bibliography{references}

\input{50_biography.tex}

\EOD

\end{document}

%% file: 10_introduction.tex
\section{Introduction}\label{sec:introduction}

\IEEEPARstart{E}{\lowercase{motion-aware}} interaction has been identified as a key factor in improving human-computer interaction. To this end, much research has been devoted to automatic speech-emotion recognition (SER). Although SER within the same corpus has shown reasonable accuracy, cross-corpus SER performance remains a challenge \cite{Schuller2010Cross-CorpusStrategies, Milner2019ARecognition, Gideon2021ImprovingADDoG}.
The ability to perform cross-corpus SER is critical to achieving emotion-aware interactions in real-world applications. This is because speech signals obtained from different devices, recording backgrounds, spoken languages, and acoustic signal conditions can differ from those in the training dataset and real-world implementation~\cite{Schuller2010Cross-CorpusStrategies}.

Researchers have investigated domain adaptation techniques to enhance cross-corpus SER performance \cite{Mao2016DomainClasses, Liu2018UnsupervisedLearning, Xi2019SpeakerAdapters, Ocquaye2021CrossNetwork}. Domain adaptation is a transfer learning method that involves training a model to optimise for a different distribution than the one used in the initial training. 
Researchers now use different types of domain adaptation strategies like, adversarial based, knowledge distillation based, and ensemble based~\cite{CS2021UnsupervisedLanguages,Park2019UnsupervisedRecognition, Meng2019DomainRecognition,Singhal2023DomainApplications}.
However, current domain adaptation approaches have a significant drawback: they cannot adapt to constantly changing real-world settings. In such settings, an intelligent agent interacting with customers/users could benefit from dynamically updating itself when it misclassifies speech emotion. 

In this study, we address the drawback of the inability to adapt to the constantly changing real-world, by proposing a domain adaptation technique based on reinforcement learning (RL)\HL{, which is uniquely suited for the dynamic requirements of real-world SER applications}. Our approach leverages the dynamic updating capability of RL to develop a cross-corpus SER technique that can adapt in real-time.

Reinforcement learning has been utilised to optimise real-world interactions with robots and machines, following training in a simulated environment \cite{Tan2018Sim-to-Real:Robots, Balaji2020DeepRacer:Learning}. However, the application of RL in the field of SER has not been explored. In a typical RL scenario, an agent takes actions in an environment, which are interpreted as a reward and a representation of the state, and fed back to the agent. This framework does not straightforwardly apply to the domain adaptation scenario of SER. As far as we are aware, our work is the first to propose and evaluate techniques for using RL in this task.

\begin{figure}[t]
        \centering
        \includegraphics[width=0.48\textwidth]{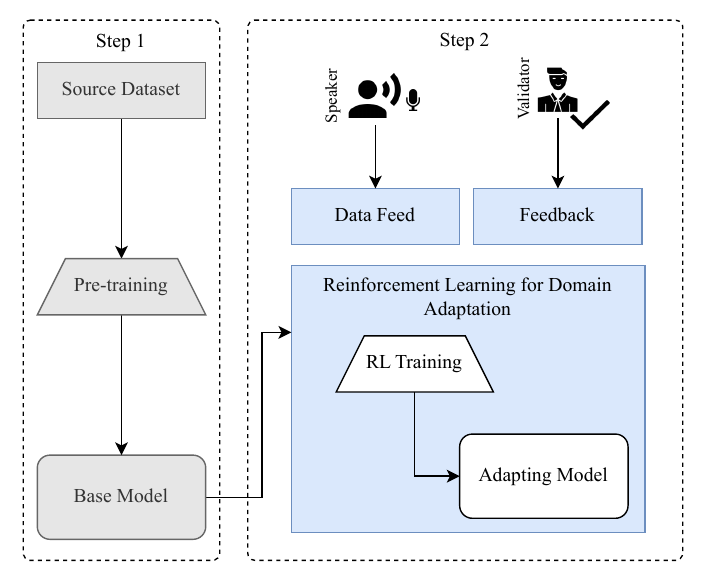}
        \caption{Overarching design for incorporating reinforcement learning in domain adaptation to enhance the accuracy of speech emotion recognition. Initially, the Base Model is pre-trained on a source dataset and subsequently optimised for the target domain using reinforcement learning, aided by user feedback and produced a Domain Adapted Model.}
        \label{fig:overall_architecture}
    \end{figure}

Figure~\ref{fig:overall_architecture} illustrates a potential use case for the proposed technique. Initially, a base deep learning model (\textit{``Base Model''}) is trained on a labelled source dataset before undergoing reinforcement learning-based domain adaptation (RL-DA). The pre-trained model parameters are then transferred to the RL agent, optimised for the target domain using RL. During this optimisation, target data are provided as states to the RL algorithm, and the user provides feedback based on the RL output. The RL-DA approach employs this feedback to calculate the loss value and optimise the RL model.
\HL{To facilitate effective domain adaptation, the RL-DA approach leverages continuous feedback from the environment. This interaction allows the RL agent to refine its emotion recognition predictions based on user or environment-provided feedback, enabling ongoing adaptation to real-world conditions without requiring manual retraining.}

\HL{The proposed methodology aims to enhance real-world adaptability of speech emotion recognition by leveraging RL for continuous self-tuning, minimising dependence on labelled data in new domains, and providing consistent, reliable emotion detection across diverse users and environments.}

Our approach leverages a RL agent to play a speech emotion recognition game. We initialise the agent's model with a pre-trained SER model, rather than random initialisation. During real-world deployment, the agent receives a speech utterance and predicts the embedded emotion. The prediction is then validated, and feedback is provided to the RL agent, serving as a reward to guide future predictions.
This framework also enables someone to deploy a working SER model in a live environment by training on a dataset available at hand rather than waiting for labelled data in the deployed environment. 

\HL{There are many practical uses for speech emotion recognition, such as improving customer service interactions}~\cite{Feng2023End-to-EndConversations}\HL{ to assisting in medical diagnosis and tracking emotional well-being}~\cite{Dhuheir2021EmotionSurvey}\HL{. SER improves customer satisfaction in customer service by allowing systems to react adaptively to a user's emotional state. It can enable physicians in managing mental health disorders by monitoring and identifying changes in patients' emotional well-being. SER models must, however, dynamically adjust to shifting conditions, such as different accents, varying noise levels, and different speaking patterns, in order to perform at their best in these real-world applications.
In contrast to classic supervised models, which necessitate retraining with fresh labelled data, self-tuning models—like those based on RL—are significant because they can continually update and improve their performance based on real-time feedback. Self-tuning models can adapt to these changes in dynamic situations offering accurate and dependable emotion detection in constantly changing real-world contexts. This feature is essential for maintaining SER applications' stability and efficacy by providing constant performance across various user groups and environments.}

\HL{In order to make models more resilient to changes in input data, multi-condition training is a technique used in machine learning, mostly in speech processing. Deep Learning models are trained on data from multiple conditions (e.g., varying noise levels, settings, or speakers)}~\cite{Parada2022PMCT:Recognition, Du2014RobustNetworks}\HL{. By subjecting the model to a range of situations throughout training, multi-condition training aims to increase the model's generalisation to unknown conditions.
This study addresses domain adaptation that is particularly applicable to the real-world SER scenario. Compared to the more general approach of multi-condition training, this can lead to more targeted performance gains. Domain adaptation methods can be designed to transfer knowledge from a source domain (e.g., a controlled environment) to a target domain (e.g., real-world conditions), potentially leading to better performance in the target domain.
While multi-condition training aims for robustness across an extensive range of conditions, domain adaptation concentrates on specific domains and may not generalise effectively to completely new or unforeseen conditions outside the target domain.}

We developed an example application to demonstrate the above use case\footnote{\url{https://rlemotion.cloud.edu.au}}. This application consumes recorded audio and infers the emotion from that audio utterance while users can provide feedback on the inferred emotion. This feedback is 
applied for RL optimisation.

We evaluate our proposed approach considering widely used speech corpora in cross-corpus and cross-language scenarios. Our results demonstrate that our model achieves better SER performance than fully supervised benchmark models with an accuracy improvement of at least 11\%. Additionally, we simulate a real-world data feed scenario and show that our RL-based model significantly outperforms the fully supervised benchmark models.
We also focus on evaluating the behaviour of our approach under speaker dependent and speaker independent settings. The results show that both speaker dependent and speaker independent settings outperform the fully supervised benchmark models.

This study focuses on, 1) Building a domain adapting speech emotion recognition model which can be used in the real-world. 2) Developing a RL based domain adapting framework which can be easily deployed in the real-word without waiting for the labelled data from the new domain.

%% file: 20_literature_review.tex
\section{Related Work}

Speech emotion recognition aims to bridge the gap between human-machine interactions by enabling systems to understand and respond to emotional cues. Traditional approaches relied on handcrafted features such as pitch, energy, and spectral properties, but recent advancements in deep learning have revolutionised SER by leveraging automatic feature extraction and robust modelling techniques~\cite{Bagadi2024ARecognition, Mishra2024ImprovementMethod, Hyeon2024ImprovingExperts, Saleem2024Squeeze-and-excitationRecognition, Saleem2023DeepCNN:Recognition, Bagadi2023AnRecognition}.

Deep Reinforcement Learning (Deep RL) is a novel approach that combines the techniques of RL and Deep Neural Networks (DNN) and has gained popularity with advancements in deep learning. Primarily, Deep RL has been implemented in gaming applications~\cite{Szita2012ReinforcementGames, Mnih2013PlayingLearning, Carr2018DomainAtari}, recommendation systems~\cite{Munemasa2018DeepSystems}, and robotics~\cite{Zhao2020Sim-to-RealSurvey}. However, the application of RL in speech-based domains is gradually gaining momentum~\cite{Latif2022AApplications}.

This section presents an overview of the existing literature in two distinct groups: one group combines the literature on Domain Adaptation in SER, while the other group consolidates the previous works that employed RL for domain adaptation.
	
\subsection{Domain Adaptation in Speech Emotion Recognition}
\label{sec:lit:domain_adaptation}

Domain adaptation was introduced to overcome the ``corpus bias'' issue. Since deep learning methods are powerful in extracting non-linear features from the input, domain adaptation can easily be implemented on deep learning-based platforms, yielding more robust and better performing models \cite{Yu2013FeatureTasks, NiiNoi2018CoupledRecognition}. 
	
	\begin{table*}
		\centering
		\caption{Summary and focus on the literature on Reinforcement Learning (RL) for Domain Adaptation (DA) and Speech Emotion Recognition (SER).}
		\label{tab:literature}
		\renewcommand{\arraystretch}{2}
		\begin{tabular}{|l|ccc|p{5.5cm}|p{4.5cm}|}
			\hline
			\multirow[c]{2}{*}{Paper} & \multicolumn{3}{c|}{Focus}  & \multirow[c]{2}{*}{Brief Experimental Set-up \& Results}  & \multirow[c]{2}{*}{Methodology}     \\ \cline{2-4} 	&
			\multicolumn{1}{l|}{RL} &
			\multicolumn{1}{l|}{DA} &
			\multicolumn{1}{l|}{SER} & 
			& \\ \hline
			Mao et al, 2016~\cite{Mao2016DomainClasses} & \multicolumn{1}{c|}{\xmark} & \multicolumn{1}{c|}{\cmark} & \multicolumn{1}{c|}{\cmark} & 61.54\% UAR for ABC~\cite{Schuller2007AudiovisualSpaces} \& 57.58\% UAR for Emo-DB~\cite{Burkhardt2005ASpeech} & Sharing priors between source \& target \\ \hline
   
			Carr et al, 2018~\cite{Carr2018DomainAtari} & \multicolumn{1}{c|}{\cmark} & \multicolumn{1}{c|}{\cmark} & \multicolumn{1}{c|}{\xmark} & Converges solution 800000 steps faster in Pong Game using  Arcade Learning Environment~\cite{Bellemare2013TheAgents} & Adversarial autoencoder aligns source and target domains via feature space  \\ \hline
   
			Lakomkin et al, 2018~\cite{Lakomkin2018EmoRL:Learning} & \multicolumn{1}{c|}{\cmark } & \multicolumn{1}{c|}{\xmark} & \multicolumn{1}{c|}{\cmark} & 84.9\% Accuracy with 1.82x speed-up for IEMOCAP~\cite{Busso2008IEMOCAP:Database} & Action-based emotion detection with GRU  \\ \hline
   
			\pbox{3cm}{Hossain and Muhammad, 2019~\cite{Hossain2019EmotionData}} & \multicolumn{1}{c|}{\xmark} & \multicolumn{1}{c|}{\xmark} & \multicolumn{1}{c|}{\cmark} & 86.4\% Accuracy for ELM fusion using eNTERFACE 05~\cite{Martin2006TheDatabase} \& 99.9\% for ELM fusion using Recorded Data & Multimodal fusion of audio-visual data using deep convolutional networks \\ \hline
   
			Koo et al, 2019~\cite{Koo2019AdversarialSystems} & \multicolumn{1}{c|}{\cmark} & \multicolumn{1}{c|}{\cmark} & \multicolumn{1}{c|}{\xmark} & Task Success Rate of 90.1\% for San Francisco Restaurant dataset to Cambridge Restaurant dataset& Adversarial calibrator aligns feature extractors across domains using RL \\ \hline
   
			Arndt et al, 2020~\cite{Arndt2020MetaAdaptation} & \multicolumn{1}{c|}{\cmark} & \multicolumn{1}{c|}{\cmark} & \multicolumn{1}{c|}{\xmark} & Policy trained with meta-learning exhibited less variance \& improved adaptation compared to domain randomisation & Meta-policy training with trajectory generation for rapid sim-to-real adaptation \\ \hline
   
			Ahn et al, 2021~\cite{Ahn2021Cross-CorpusAdaptation} & \multicolumn{1}{c|}{\xmark} & \multicolumn{1}{c|}{\cmark} & \multicolumn{1}{c|}{\cmark} & 50.8\% UA for datasets MSP-IMPROV~\cite{Busso2017MSP-IMPROV:Perception}, EMO-DB~\cite{Burkhardt2005ASpeech}, and KME-Korean) using IEMOCAP as source & Few-shot learning with DA for cross-corpus emotion classification \\ \hline
   
			Ishaq et al, 2023~\cite{Ishaq2023TC-Net:Network} & \multicolumn{1}{c|}{\xmark} & \multicolumn{1}{c|}{\xmark} & \multicolumn{1}{c|}{\cmark} & 78.34\% UA for IEMOCAP \& 91.61\% UA for EMO-DB & Extract temporal speech features using TCN \& classify emotions via FCN \\ \hline
   
			Khan et al, 2024~\cite{Khan2024MSER:Fusion} & \multicolumn{1}{c|}{\xmark} & \multicolumn{1}{c|}{\xmark} & \multicolumn{1}{c|}{\cmark} & 72.30\% UA for IEMOCAP and MELD~\cite{Poria2018MELD:Conversations} & Cross-attention, deep fusion, CNNs \& speech-text alignment based SER \\ \hline
   
			\textbf{This paper} & \multicolumn{1}{c|}{\cmark} & \multicolumn{1}{c|}{\cmark} & \multicolumn{1}{c|}{\cmark} & UAR\% is measured for datasets: MSP-IMPROV, IEMOCAP, ESD~\cite{Zhou2022EmotionalESD}, and EmoDB. Results shown in Section~\ref{sec:results} & Pre-train model, RL optimises DA using feedback \\ \hline
			
		\end{tabular}%
		\renewcommand{\arraystretch}{1.0}
	\end{table*}

Mao et al.~\cite{Mao2016DomainClasses} conducted experiments on domain adaptation using a two-class task, where they shared priors between the source and target domains. They pre-trained a two-layer neural network using unsupervised learning and shared the standard classifier parameters between the source and target domains. In contrast, our approach follows the RL paradigm for domain adaptation and involves pre-training the source model with a labelled source dataset. However, we cannot compare our results with Mao et al.'s study on SER and domain adaptation since they used a different dataset and a different number of classes for prediction.

Gharib et al.~\cite{Gharib2018DetectionEvents} proposed unsupervised adversarial domain adaptation and pre-trained the model with two conditional sets, resulting in a nearly 10\% increase in accuracy. Ahn et al.~\cite{Ahn2021Cross-CorpusAdaptation} proposed a few-shot learning methodology in unsupervised domain adaptation for cross-corpus SER tasks, where they used multiple corpora to optimise the emotion recognition robustness to unseen samples.

Abdelwahab and Busso~\cite{Abdelwahab2015SupervisedSpeech} studied the best adaptation technique for speech emotion recognition models through supervised domain adaptation or online training. They also investigated how the size of the labelled data affects the performance of the resulting model.

Adversarial Domain Adaptation is a stream of research in domain adaptation~\cite{Yin2020Speaker-InvariantRecognition, Mathur2020UnsupervisedClassification, Lu2022DomainRecognition}. Latif et al.~\cite{Latif2019UnsupervisedRecognition} proposed using adversarial training techniques to improve cross-lingual domain adaptation. This unsupervised learning method employs two distinct auto-encoders for the source and target and a single discriminator to determine whether an audio utterance is fake or real. Although we focus on cross-lingual domain adaptation, our main objective is enabling domain adaptation in real-world applications.

\HL{Zhang et al.}~\cite{Zhang2022UnsupervisedRecognition}\HL{ introduces a novel approach for cross-corpus SER using unsupervised domain adaptation, combining transformers with mutual information maximisation to align feature distributions across domains effectively. This approach demonstrates significant improvements in emotion recognition accuracy across multiple cross-corpus benchmarks, outperforming several baseline models and showcasing enhanced adaptability in SER tasks using IEMOCAP and MSP-Improv datasets.}

\HL{Mote et al.}~\cite{Mote2024UnsupervisedConversion}\HL{ introduces unsupervised domain adaptation technique which involves a k-nearest neighbour-based voice conversion technique to adapt unlabelled speech data for emotion recognition by transforming it to match the labelled data domain. Their methodology avoids model re-training requirements for the unlabelled data. Results obtained using MSP-Improv and MSP-Podcast datasets show an 8.2\% performance increment for valence detection over their baseline.} 
	
\subsection{Reinforcement Learning for Domain Adaptation}
\label{sec:lit:rl_for_da}
The RL agent acquires knowledge through experiences gained by exploring and exploiting the environment. The agent selects the best action based on the policies, known as exploitation, given the current state. However, in the initial phase, the agent lacks sufficient experience to update the policy, so it explores the environment by randomly selecting actions and storing the experience in the memory buffer with a reward. Exploration poses a challenge since some actions in certain states may be unsafe~\cite{Arndt2020MetaAdaptation}.

Prior knowledge can be provided to the RL agent in various ways, such as demonstration-based learning \cite{Nair2018OvercomingDemonstrations, Zhu2020TransferSurvey}, pre-training, and domain adaptation \cite{Arndt2020MetaAdaptation}. These methods reduce the exploration time and the risk of unsafe exploration. Previous research has demonstrated that pre-training reduces the training time while achieving higher performance \cite{Rajapakshe2019Pre-trainingRecognition, Rajapakshe2020DeepRecognition, Rajapakshe2022ARecognition}.

Koo et al.~\cite{Koo2019AdversarialSystems} proposed using adversarial training to enhance the coherent training of the target domain feature extractor, which is considered a generator for the target domain feature extractor. The parameters are similar to the latent feature set in the source and are trained with an adversarial training loss.


Hazara and Kyrki~\cite{Hazara2019TransferringWorld} proposed a transfer approach that captures the core features of a simulated system and rationalises the dynamics combined with incremental learning. They demonstrated their approach in a basketball task with a robot and showed that the target model has improved task generalisation capability compared to direct usage.

While these previous studies have employed an RL-based methodology for domain adaptation, none have investigated speech emotion recognition.

{\noindent\bf In summary:}
\begin{enumerate}
\item The extensive literature on domain adaptation for SER employs various deep learning techniques.
\item While RL has been used in a few studies, including EmoRL \cite{Lakomkin2018EmoRL:Learning}, for speech emotion recognition tasks, no research has yet explored the application of RL for domain adaptation in SER.
\item Table~\ref{tab:literature} provides a brief overview of the literature on RL for domain adaptation and speech emotion classification, highlighting the absence of research that combines RL with domain adaptation for SER, which is the primary focus of this paper.
\end{enumerate}

%% file: 30_methodology.tex
\section{Methodology}
In this section, we outline our approach for domain adaptation using RL for SER. We first provide a brief introduction to RL, followed by a description of our proposed approach and the baselines used to compare its performance.

\subsection{Background}
\begin{figure*}[ht]
\centering
\includegraphics[width=0.9\textwidth]{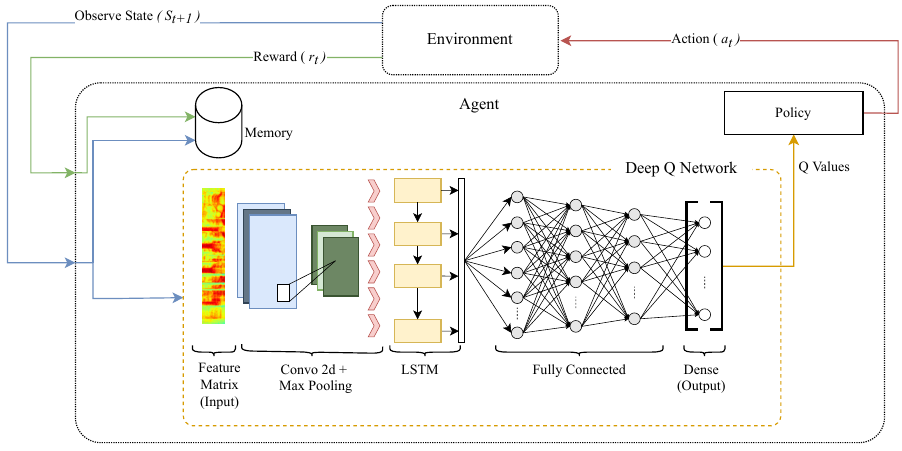}
\caption{The architecture of a Reinforcement Learning system featuring a Deep Q Network. The RL Agent comprises several constituent parts: Memory, RL policy, and DQN. Interacting with a simulated environment, the Agent receives emotional cues (actions) input and produces an output in response, which comprises feedback and the subsequent audio utterance (state).}

\label{fig:rl_architecture_model_recipe}
\end{figure*}

Reinforcement Learning is a machine learning paradigm that draws inspiration from behaviourist psychology and mimics the learning process of a child acquiring new skills. An RL problem typically comprises two fundamental components, namely the Environment and the Agent. In this context, the Agent interacts with the Environment by performing actions, and the Environment responds by providing a reward that corresponds to the action taken along with the updated state of the Environment. Figure \ref{fig:rl_architecture_model_recipe} depicts the general architecture of the RL methodology in a concise manner.

A Markov Decision Process (MDP) is a formal framework that can be used to model an RL problem, which comprises of state space $S$, action space $A$, and reward $R$. A state $s \in S$ represents the current state of the environment, action $a \in A$ is a single task that can be performed on the environment, and reward $r \in R$ is the feedback value returned from the environment after executing the task $a$ for the state $s$. The RL agent learns a policy $\pi(s,a)$ that represents the probability of selecting action $a$ for a given state $s$. The Q-value, also known as the quality value, represents the expected reward when action $a$ is performed on an environment with state $s$. The discount factor $\gamma$ is used to weigh the immediate reward against the expected long-term reward. The optimum Q-value $Q^*(s,a)$ can be written as in Equation~\ref{eq:bellmans_state_action_value_function};
	\begin{equation}
		\label{eq:bellmans_state_action_value_function}
		Q^*_\pi(s,a) =max \mathbb{E}_\pi[R_{t+1} + \gamma q_\pi(S_{t+1},A_{t+1})].
	\end{equation}
	Q-values are estimated using Q-tables in q-learning, while a deep neural network called Deep Q Network (DQN) is used to estimate the Q values in deep q-learning. 
	The loss function in Equation~\ref{eq:loss_function} is used to calculate the loss and $Q_{target}$ can be calculated using Equation~\ref{eq:q_target}.
	\begin{equation}
		\label{eq:loss_function}
		L = \Sigma{(Q_{target} - Q)^2}, 
	\end{equation}
	\begin{equation}
		\label{eq:q_target}
		Q_{target} = R(s_{t+1},a_{t+1}) + \gamma Q(s_{t+1},a_{t+1}).
	\end{equation}
	Combining the two equations~\ref{eq:loss_function} and \ref{eq:q_target}, the loss function can be rewritten as in Equation~\ref{eq:final_loss_function}.
	\begin{equation}
		\label{eq:final_loss_function}
		L = \Sigma{(R(s_{t+1},a_{t+1}) + \gamma Q(s_{t+1},a_{t+1})  - Q(s_t,a_t) )^2}.
	\end{equation}
	The loss $L$ is minimised in the training phase of RL. 

\subsection{Domain Adapting Reinforcement Learning for Speech Emotion Recognition }
This study employs RL as a game for recognising emotions. The RL agent aims to predict the correct emotion (action $a$, \textit{in RL terms}) for a given audio utterance (state $s$, \textit{in RL terms}). The environment returns a reward $r$, obtained 
through feedback. The RL agent learns a policy $\pi$ to maximise the reward obtained at each episode. The state space $S$ is defined as the distribution of speech audio, the action space $A$ as the discrete emotion classes, and action selection involves inferring an emotion from a given audio utterance. According to Equation~\ref{eq:final_loss_function}, minimising $L$ requires maximising $Q(s_t,a_t)$ while also maximising the reward $R_{t+1}$ at optimum Loss ($L^*$).

\subsubsection{Selection of Reinforcement Algorithm}
The implementation of RL can be achieved through various algorithms, each characterised by unique properties. When selecting an RL algorithm, it is essential to consider the type of state space (discrete or continuous) and action space (discrete or continuous). Our SER problem is defined by a discrete state space and discrete action space, making Q-Learning-based algorithms an appropriate choice~\cite{Wang2011ReinforcementApplication}. We therefore utilise Deep Q-Learning, which exploits the capabilities of Deep Neural Networks to approximate Q-values from states, enabling efficient learning in complex environments.

\subsubsection{Pre-training Before RL}
Before using an RL agent, pre-training the DQN model has been shown to enhance its performance \cite{Rajapakshe2020DeepRecognition, Rajapakshe2022ARecognition}. This study uses RL to optimise the pre-trained model for the target domain. The RL agent with a pre-trained DQN model interacts with the environment and attempts to adapt the DQN to the target domain.

\subsection{Baseline}

RL methodologies typically do not report performance in terms of accuracy, due to the absence of labelled data in RL experiments. Therefore, we employed a supervised learning (SL) approach as a baseline to facilitate a comparative evaluation of our methodology's performance.
To this end, we adopt the DNN architecture defined in the DQN of the RL agent (as depicted in Figure~\ref{fig:rl_architecture_model_recipe}) as the architecture for the SL approach. First, we train the model using the Source Dataset to align with the pre-training in the RL approach, resulting in a Base Model as shown in Figure~\ref{fig:overall_architecture}. Next, the Base Model is trained using the Target dataset to match the RL-DA. This enables us to compare our RL-DA and Supervised Learning-based Domain Adaptation using similar parameters.

To evaluate the performance of our proposed RL-DA approach, we utilise four labelled datasets, with 20\% of each dataset reserved for testing, while the remainder is used for either pre-training or RL domain optimisation. We provide more details about the used datasets in Section~\ref{sec:datasets}, while Section~\ref{sec:results} outlines the various types of experiments conducted.

%% file: 35_experimental_setup.tex
\section{Experimental Setup}
\label{sec:experimental_setup}

In this section, we elaborate on the datasets, input features, and model configurations. Figure~\ref{fig:rl_architecture_model_recipe} depicts the detailed RL architecture and the DQN model employed in our study.

To simulate the RL environment, we receive the emotion (action) from the RL agent and output the reward and subsequent audio utterance as the state. The environment determines the reward by comparing the inferred class (agent's action) with the ground truth label in the dataset.\HL{ Reward is positive if the action is inferred correctly and negative otherwise.}
In commercial or real-world applications, this environment can be replaced by integrating it into a feedback system\HL{ where the feedback is generated considering the interaction of the user. }. 
\HL{The convergence criteria is measured using the loss difference between the inferred and target Q-Values. The model is considered converged, when the loss difference is less than 0.2 for 100 iterations.}

Each feedback from the environment is stored in a memory database along with the audio utterance (state), reward, and inferred emotion (action), which are then utilised to optimise the DQN model after a specified number of iterations. The DQN model is a Deep Neural Network that comprises a combination of Convolutional Neural Network (CNN), Long Short-Term Memory (LSTM), and dense layers. It ingests the Mel Frequency Cepstral Coefficients (MFCC) feature matrix and outputs the estimated Q-values required for the RL policy to determine the emotion (action). Therefore, the DQN model estimates the Q-values for a given audio utterance (state), and these Q-values are used by the RL policy to determine the emotion (action).

\subsection{Datasets}
\label{sec:datasets}
We utilise four publicly available datasets widely used in the field of SER: MSP-IMPROV \cite{Busso2017MSP-IMPROV:Perception}, IEMOCAP \cite{Busso2008IEMOCAP:Database}, ESD \cite{Zhou2022EmotionalESD}, and EmoDB \cite{Burkhardt2005ASpeech}. These datasets enable the use of Cross-Corpus ($CC$) and Cross-Language ($CL$) experiments.

Due to the unbalanced number of utterances under each emotion class for all datasets, we use an audio augmentation technique to generate audio utterances. We use Vocal Tract Length Perturbation (VLTP) \cite{Jaitly2013VocalRecognition} to augment audio utterances, which balances the number of utterances under each emotion class when preparing the testing subsets. All four speech emotion datasets utilised in this study contain categorical emotional labels, and we selected four emotions widely used in the literature: happiness, sadness, anger, and neutral.

Additionally, we use the DEMAND~\cite{Thiemann2013DEMAND:Environments} dataset's kitchen environment audio as the background sound to evaluate the proposed
RL-DA approach in real-world scenarios. Section~\ref{sec:eval_exp_3} describes the usage of the DEMAND dataset in the experiments. Below, we provide more details on the considered datasets. 

\subsubsection{IEMOCAP (IEM)}
IEMOCAP is a widely used dataset that comprises 12 hours of acted multi-speaker audio-visual data. The IEMOCAP dataset consists of dyadic sessions featuring both scripted and improvised scenarios. It contains categorical emotion labels of happiness, sadness, anger, and neutrality, as well as dimensional labels of valence, dominance, and activation \cite{Busso2008IEMOCAP:Database}. In this study, we utilise the audio data modality from the improvised scenarios and restrict our analysis to the categorical emotion labels.

\subsubsection{MSP-IMPROV (MSP)}
MSP-IMPROV is an audio-visual database comprising acted performances that is widely used in multi-modal speech emotion recognition research. While originally designed for an audio-visual emotional perception study, it has also been utilised in various speech emotion recognition studies \cite{Peng2020SpeechFront-Ends, Xie2021SpeechInput, Ahn2021Cross-CorpusAdaptation, Gao2022Domain-InvariantRecognition}, rendering it a suitable dataset for our research purposes. The database was recorded in a controlled environment, with 20 pre-determined scripts that encompass the primary emotions of happiness, sadness, anger, and neutrality \cite{Busso2017MSP-IMPROV:Perception}.

\subsubsection{Emotional Speech Dataset (ESD)}
The Emotional Speech Database (ESD) is a publicly available database of speech data originally designed for the purposes of speech synthesis and voice conversion. The database includes utterances from 20 speakers, consisting of 10 Mandarin and 10 English speakers, and each utterance has been categorised into one of five emotion classes: happy, surprise, neutral, angry, and sad \cite{Zhou2022EmotionalESD}. For our study, we exclusively selected utterances from 10 English speakers that correspond to the emotions of happiness, sadness, anger, and neutrality. 

\subsubsection{Berlin EmoDB (EmoDB)}
EmoDB is a database of emotional speech utterances in the German language that features recordings from 10 actors (5 male and 5 female) between the ages of 21 and 35. Each actor recorded 10 scripted texts, covering 7 different emotions, including anger, boredom, disgust, fear, happiness, sadness, and neutrality \cite{Burkhardt2005ASpeech}. In our cross-language experimental analysis, we selected EmoDB as the target dataset and focused solely on the utterances from the four emotion classes of anger, happiness, sadness, and neutrality. 

\subsubsection{Diverse Environments Multichannel Acoustic Noise Database (DEMAND)}
The DEMAND dataset is a widely utilised collection of real-world background noises. This dataset includes background noises from 18 different environments, categorised into six categories. The audio recordings from each environment were captured using an array of 16 microphones and stored in 16 channels. The DEMAND dataset offers versions of both 16 kHz and 48 kHz sampling rates; in this study, we employed the 48 kHz version and down-sampled it to 22 kHz to align with the other datasets used in our analysis.

A summary of the datasets employed in this study is presented in Table~\ref{tab:dataset_summary}.

\begin{table}[ht]
    \caption{Summary of the datasets used in this study. The `\# Classes` column contains the total number of classes in the original datasets irrespective of the four emotions (`happiness', `sadness', `anger', and `neutral') we used in our study.}
    \label{tab:dataset_summary}
    \centering
    \renewcommand{\arraystretch}{1.2}
    \begin{tabular}{l|l|l|l|l}
\multicolumn{1}{c|}{\textbf{Dataset}} & \multicolumn{1}{c|}{\textbf{\# Training}} & \multicolumn{1}{c|}{\textbf{\# Testing}} & \multicolumn{1}{c|}{\textbf{\# Classes}} & \multicolumn{1}{c}{\textbf{Language}} \\ \hline
IEMOCAP    & 2780        & 852        & 4                  & English  \\
MSP-Improv & 432         & 148        & 4                  & English  \\
ESD        & 7744        & 2428       & 5                  & English  \\
EmoDB      & 344         & 100        & 7                  & German   \\ \hline
\end{tabular}
\renewcommand{\arraystretch}{1.0}
    \end{table}

\subsection{Input features}
\label{lab:input_features}
In this study, we employed MFCC as the input features for our analysis congruent to previous studies~\cite{Likitha2017SpeechMFCC, Latif2019DirectSpeech, Patni2021SpeechFeatures,Dolka2021SpeechFeatures}. Specifically, we set the frame length to $2,048$ and the hop length to $512$, and extracted $40$ MFCCs using the Librosa python library for audio and music analysis \cite{McFee2015Librosa:Python}.
    
\subsection{Model configuration}
In this study, we employ a combined model of CNN and LSTM as it facilitates learning both frequency and temporal components in the speech signal \cite{Latif2020DeepTrends, Rajapakshe2022ARecognition}. The discriminative features of the model are learnt by stacking CNN, LSTM, and fully connected layers, respectively, as depicted in Figure~\ref{fig:rl_architecture_model_recipe}. The feature matrix is processed through two layers of 2D convolution with filter sizes of $5$ and $3$. The output of the first 2D convolution layer is batch-normalised, and a hyperparameter batch size of $128$ is used.

The output from the second 2D convolution layer is flattened and fed into an LSTM layer with $16$ cells, followed by a fully connected layer with $256$ units. A dropout rate of $0.3$ is applied before the last Dense output layer. The number of units in the output layer is set to four, which is the number of emotions to be classified. The linear activation function of the output layer is used in the RL agent as it outputs Q-values for a specific state, which should not be normalised. The input shape of the model is $40 \times 87$, where $40$ MFCCs are used in the input, and $87$ is the number of MFCC frames.

The model is optimised using the Adam Optimiser with a $2.5\times10^{-4}$ learning rate. The Deep Learning API Keras \cite{Chollet2015Keras} with Tensorflow \cite{Abadi2015TensorFlow:Systems} (version 2.1.0) is used as the back-end for modelling and training purposes in this study. 

The model consists of 42,966 trainable parameters and 87,327 floating-point operations (FLOPs) which gives a perspective on the complexity of the model. Considering the FLOPs, it indicates that the model is lightweight, making it scalable and suitable for deployment in resource-constrained environments as well. Low parameter count leads to a faster convergence and reduced memory requirements.

%% file: 40_results.tex
\section{Evaluation}
\label{sec:results}
We devised and executed several experiments to evaluate the performance of the proposed RL-DA approach. We formulated three scenarios: (1) pre-training with a source dataset and domain optimisation with a target dataset separately, (2) pre-training with a subset of the source dataset and domain optimisation with a mix of the remaining source and target datasets, and (3) domain adaptation in a recreated real-world setting. Each scenario's specifics and findings are outlined in Sections~\ref{sec:eval_exp_1}, \ref{sec:eval_exp_2}, and~\ref{sec:eval_exp_3}, respectively. We also evaluated the proposed RL-DA method for speaker-independent SER in Section~\ref{sec:eval_exp_4}.

\subsubsection{Selection of Source and Target Datasets}
Since deep RL algorithms learn from feedback instead of directly labelled data, they necessitate more data instances than supervised learning algorithms~\cite{Mnih2015Human-levelLearning, Sutton2018ReinforcementIntroduction}. Therefore, we opted for IEMOCAP and MSP-Improv as the source datasets and ESD and EmoDB as the target datasets in cross-corpus and cross-language experiments, respectively. In the RL agent, the Max Boltzman Policy (Max.B)~\cite{Wiering1999ExplorationsLearning} serves as the RL policy.

\subsubsection{Evaluation Metrics}
We use the Unweighted Average Recall Rate (UAR,\%), which is a well-known metric researchers use to assess the performance of speech-based machine learning models \cite{Mao2016DomainClasses, Ahn2021Cross-CorpusAdaptation, Ishaq2023TC-Net:Network, Khan2024MSER:Fusion}. The recall is computed for each classification problem label, and the UAR is calculated as the unweighted average of each label's recall values. 

The Python code repository for the experiments and data preparation is available in a public GitHub repository called ``RL-DomainAdaptation''\footnote{\url{https://github.com/iot-health/RL-DomainAdaptation}}.

\subsection{Experiment with separate datasets for source and target}
\label{sec:eval_exp_1}
\begin{figure*}[h]
        \centering
        \includegraphics[width=0.75\textwidth]{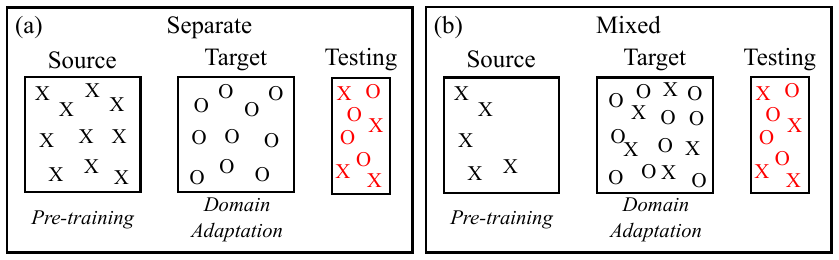}
        \caption{Composition of datasets used in Target and Source: (a) the Target dataset is not mixed with Source data and (b) the Target dataset is mixed with Source data. The source dataset is used to pre-train the base model while the target dataset is used for domain adaptation.}
        \label{fig:dataset_mixing}
    \end{figure*}

Here, we evaluate the proposed RL-DA approach for separate source and target datasets. We do this for both the cross-corpus and cross-language schemas. Table~\ref{tab:datasets_in_exp1} and Figure~\ref{fig:dataset_mixing} (a) show the datasets used for each schema. 
    \begin{table}[!ht]
    \centering
    \caption{Summary of the datasets used in the cross-corpus (CC) and cross-language (CL) experiments. (The Target dataset is not mixed with Source data)}
    \renewcommand{\arraystretch}{1.2}
    \label{tab:datasets_in_exp1}
        \begin{tabular}{l|ll|l}
        \cline{2-4}
        & \multicolumn{2}{c|}{\textbf{Training Dataset}}                                                               & \multirow{2}{*}{\textbf{Testing Dataset}} \\
        & \multicolumn{1}{c|}{\textbf{Source}} & \multicolumn{1}{c|}{\textbf{Target}} &                          \\ \hline
        \multirow{2}{*}{CC} & \multicolumn{1}{l|}{IEM}                 & ESD                                          & IEM + ESD            \\ \cline{2-4} 
        & \multicolumn{1}{l|}{MSP}              & ESD                                          & MSP + ESD         \\ \hline
        \multirow{2}{*}{CL} & \multicolumn{1}{l|}{IEM}                 & EmoDB                                        & IEM + EmoDB          \\ \cline{2-4} 
        & \multicolumn{1}{l|}{MSP}              & EmoDB                                        & MSP + EmoDB      \\ \hline
        \end{tabular}
        \renewcommand{\arraystretch}{1.0}
    \end{table}    

Initially, the model is pre-trained using the training subset of the source dataset, resulting in the \textit{``Base model''}. Subsequently, the pre-trained parameters of the Base model are transferred to the DQN of the RL agent. The RL deep Q learning algorithm is then executed to optimise the DQN model using the target dataset in the environment, leading to the domain-optimised DNN model. We employ a Supervised Learning approach to evaluate the domain-optimised model's performance. Specifically, we utilise the testing subset of the Target dataset to infer the emotion of each testing utterance and record both the inferred emotion and the labelled emotion for the utterance. We calculate the UAR value for each experiment based on the recorded data about inferred emotions and ground truth.

We also measure the performance of our methodology using a Supervised Learning approach as the baseline (SL-DA - sep). We develop a DNN model with an identical architecture to the one used in DQN (as depicted in Figure~\ref{fig:rl_architecture_model_recipe}), which is pre-trained using the Source dataset. Next, the same model is trained using the training subset of the Target dataset. Finally, the resulting model is tested using the testing subset of the Target dataset to obtain the testing accuracy.

We report the UAR of the baseline (SL-DA - sep) model after training with Supervised Learning and the UAR of the model after training with the RL-DA approach (RL-DA) in Table~\ref{tab:baselineVsRL}.

Our findings indicate that the RL-DA approach outperforms the baseline UAR in most scenarios.

\begin{table}[!ht]
    \fontsize{7.5pt}{10}\selectfont
    \caption{Comparison of performance between the Supervised Learning based (SL-DA - sep), the Reinforcement Learning based Domain Adaptation approach (RL-DA), and UAR (\%) presented in the SER Literature in a similar setup (same source and target datasets) for the cross-corpus (CC) and cross-language (CL) experiments, where the Target dataset was not mixed with Source data}
    \label{tab:baselineVsRL}
    \centering
    \renewcommand{\arraystretch}{1.2}

    \begin{tabular}{c|ll|lll}
    \cline{2-6}
    \multicolumn{1}{l|}{} & \multicolumn{2}{c|}{\textbf{Dataset}}                                       & \multicolumn{3}{c}{\textbf{UAR   (\%)}}                                                                                           \\
    \multicolumn{1}{l|}{} & \multicolumn{1}{c|}{\textbf{Source}} & \multicolumn{1}{c|}{\textbf{Target}} & \multicolumn{1}{c|}{\textbf{SL-DA - sep}}  & \multicolumn{1}{c|}{\textbf{RL-DA}}        & \multicolumn{1}{c}{\textbf{Literature}} \\ \hline
    \multirow{2}{*}{CC}   & \multicolumn{1}{l|}{IEM}             & ESD                                  & \multicolumn{1}{l|}{63.54 $\pm$ 2.11}          & \multicolumn{1}{l|}{\textbf{63.99 $\pm$ 1.95}} & -                                       \\
                          & \multicolumn{1}{l|}{MSP}             & ESD                                  & \multicolumn{1}{l|}{63.18 $\pm$ 3.34}          & \multicolumn{1}{l|}{\textbf{66.10 $\pm$ 0.84}} & -                                       \\ \hline
    \multirow{2}{*}{CL}   & \multicolumn{1}{l|}{IEM}             & EmoDB                                & \multicolumn{1}{l|}{\textbf{74.67 $\pm$ 1.03}} & \multicolumn{1}{l|}{73.17 $\pm$ 0.62}          & 62.91 {[}53{]}                          \\
                          & \multicolumn{1}{l|}{MSP}             & EmoDB                                & \multicolumn{1}{l|}{56.67 $\pm$ 8.26}          & \multicolumn{1}{l|}{\textbf{65.50 $\pm$ 0.71}} & -                                       \\ \hline
    \end{tabular}
    \renewcommand{\arraystretch}{1.0}
    \end{table}
Furthermore, when comparing the SL-DA method and the RL-DA method using the results in Table~\ref{tab:baselineVsRL}. We observe that the standard deviation of the cross-language schema has decreased from 8.26 to 0.71. This demonstrates that the model performances are more consistent using the RL-DA approach.
This outcome is significant because, unlike SL approaches that rely on labelled data to train the model, RL methods only provide feedback rewards indicating the accuracy of inference made during the RL optimisation phase. Therefore, achieving SL accuracy with the RL-DA approach implies that the method has successfully trained the model to the SL standard.

\subsection{Experiment mixing the source and target datasets}
\label{sec:eval_exp_2}
This subsection assesses the effectiveness of our proposed methodology, where a subset of the source dataset is included in the RL optimisation phase to preserve the generalisable features of the source dataset.

We conduct this experiment similar to the previous experiment mentioned in Section~\ref{sec:eval_exp_1}, with the following modifications: only a subset of the training set of the source dataset is employed in the pre-training phase. In contrast, a combination of the remaining training set of the source dataset and the training set of the target dataset is utilised in the RL optimisation phase. Table~\ref{tab:datasets_in_exp2} and Figure~\ref{fig:dataset_mixing} (b) depict how the datasets are employed in this experiment. We employ SL as a baseline (SL-DA - mix) and measure the performance to compare our proposed approach.
    
    \begin{table}[!ht]
    \centering
    \caption{Composition of datasets used in Target and Source when experimenting by mixing datasets in the cross-corpus (CC) and cross-language (CL) schemata}
    \label{tab:datasets_in_exp2}
    \renewcommand{\arraystretch}{1.2}
        \begin{tabular}{l|ll|l}
        
        \cline{2-4}
        & \multicolumn{2}{c|}{\textbf{Training Dataset}}        & \multirow{2}{*}{\textbf{Testing Dataset}} \\
        & \multicolumn{1}{c|}{\textbf{Source}} & \multicolumn{1}{c|}{\textbf{Target}} &                \\ \hline
        \multirow{2}{*}{CC} & \multicolumn{1}{l|}{50\% IEM}     & 50\% IEM + ESD    & IEM + ESD        \\ \cline{2-4} 
                            & \multicolumn{1}{l|}{50\% MSP}     & 50\% MSP + ESD    & MSP + ESD        \\ \hline
        \multirow{2}{*}{CL} & \multicolumn{1}{l|}{50\% IEM}     & 50\% IEM + EmoDB  & IEM + EmoDB      \\ \cline{2-4} 
                            & \multicolumn{1}{l|}{50\% MSP}     & 50\% MSP + EmoDB  & MSP + EmoDB      \\ \hline
        \end{tabular}
    \renewcommand{\arraystretch}{1.0}
    \end{table}

    Results of the experiments done by mixing source data into the target dataset as mentioned above are presented in Table~\ref{tab:mixing}. The accuracy of the RL-DA approach is compared with the Supervised Learning approach -- SL-DA - mix. 
   
    \begin{table}[ht]
    \caption{Comparison of performance between Supervised Learning based (SL-DA - mix) and Reinforcement Learning based Domain Adaptation approach (RL-DA) in the experiment with mixed target datasets for cross-corpus (CC) and cross-language (CL) schemata.}
    \label{tab:mixing}
    \centering
    \renewcommand{\arraystretch}{1.2}
    \begin{tabular}{lll|rr}
    \cline{2-5}
    &\multicolumn{2}{|c|}{\textbf{Dataset}}  & \multicolumn{2}{c}{\textbf{UAR(\%)}} \\
    &\multicolumn{1}{|c|}{\textbf{Source}} & \multicolumn{1}{c|}{\textbf{Target}}  & \multicolumn{1}{c|}{\textbf{SL-DA - mix}}  & \multicolumn{1}{c}{\textbf{RL-DA}} \\ \hline
    \multirow{2}{*}{CC} &\multicolumn{1}{|l|}{IEM}      & IEM + ESD         & \multicolumn{1}{r|}{56.22 $\pm$ 1.15}   & \multicolumn{1}{r}{\textbf{77.86 $\pm$ 0.43}} \\
                        &\multicolumn{1}{|l|}{MSP}      & MSP + ESD         & \multicolumn{1}{r|}{55.52 $\pm$ 0.97}   & \multicolumn{1}{r}{\textbf{63.51 $\pm$ 2.40}} \\ \hline
    \multirow{2}{*}{CL} &\multicolumn{1}{|l|}{IEM}      & IEM + EmoDB       & \multicolumn{1}{r|}{61.33 $\pm$ 5.44}   & \multicolumn{1}{r}{\textbf{85.67 $\pm$ 2.72}} \\
                        &\multicolumn{1}{|l|}{MSP}      & MSP + EmoDB       & \multicolumn{1}{r|}{59.17 $\pm$ 1.43}   & \multicolumn{1}{r}{\textbf{77.33 $\pm$ 1.43}}\\ \hline
    
    \end{tabular}
    \renewcommand{\arraystretch}{1.0}
    \end{table}

The RL-DA approach demonstrates superior performance over SL-DA - mix by at least 8\%, in every experiment. Since the target dataset includes elements from the source dataset, the model representations learned during the pre-training phase are maintained while the model adapts to the new domain.

\begin{figure}[h]
\centering
\includegraphics[width=0.45\textwidth]{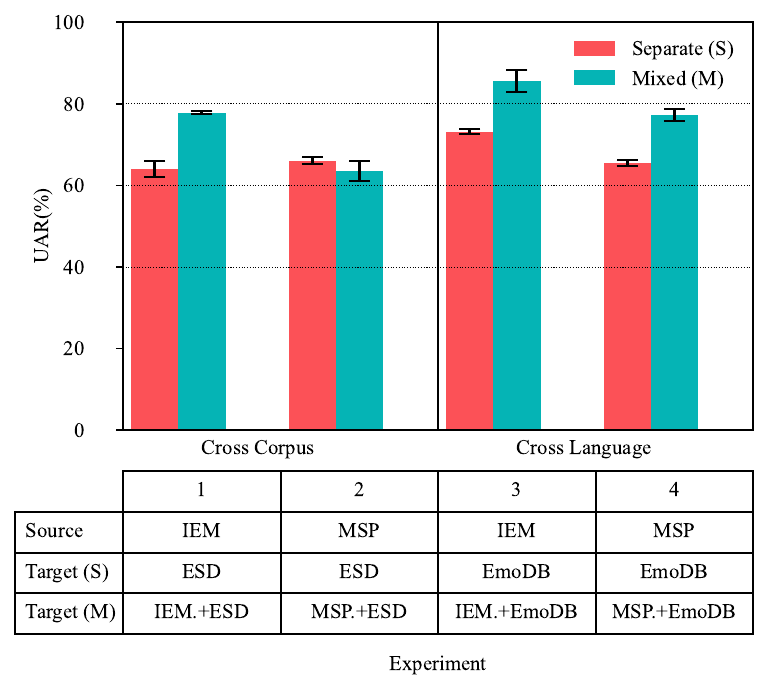}
\caption{Comparison of accuracy of each RL-DA experiment with separate datasets vs mixed datasets}
\label{fig:exp_pure_vs_mixed}
\end{figure}

We compare the performance of RL-DA in two scenarios: where the target dataset is isolated from the source dataset (S) (as discussed in Section~\ref{sec:eval_exp_1}), and where the target dataset is mixed (M) with utterances from the source dataset (as discussed in Section~\ref{sec:eval_exp_2}). These comparisons are shown in Figure~\ref{fig:exp_pure_vs_mixed}.

From the above visualisation, we note that the mixing of source and target datasets yields better performance than experiments with separate datasets. The mixing of datasets increases the diversity of the training utterances which leads the model to learn generalised feature representations when training.

\begin{table*}[ht]
\caption{Testing Accuracy of SL-DA - rw and the proposed RL-DA approach of each experiment mimicking a  real-world scenario for each cross-corpus (CC) and cross-language (CL) schema with (w/) and without (w/o) background noise. Background Noise is added to simulate real-world conditions.}
\label{tab:baselineVsRealworld}
\centering
\renewcommand{\arraystretch}{1.2}
\begin{tabular}{ll|ll|lll}
\cline{3-7}
                    & & \multicolumn{2}{c|}{\textbf{Dataset}}                                       & \multicolumn{3}{c}{\textbf{UAR(\%)}}                                                                                                        \\
                    & & \multicolumn{1}{c|}{\textbf{Source}} & \multicolumn{1}{c|}{\textbf{Target}} & \multicolumn{1}{c|}{\textbf{SL-DA - rw}}             & \multicolumn{1}{c|}{\textbf{RL-DA}}                & \multicolumn{1}{c}{$\mu_\Delta$} \\ \hline
\multirow{4}{*}{w/o noise} & \multirow{2}{*}{CC} & \multicolumn{1}{l|}{IEM} & ESD              & \multicolumn{1}{l|}{53.11 $\pm$ 1.45} & \multicolumn{1}{l|}{\textbf{62.89 $\pm$ 0.51}} &  9.78  \\
                    & & \multicolumn{1}{l|}{MSP} & ESD              & \multicolumn{1}{l|}{47.07 $\pm$ 2.56} & \multicolumn{1}{l|}{\textbf{63.18 $\pm$ 2.15}} & 16.11   \\ \cline{2-7}
& \multirow{2}{*}{CL} & \multicolumn{1}{l|}{IEM} & EmoDB            & \multicolumn{1}{l|}{60.67 $\pm$ 1.70} & \multicolumn{1}{l|}{\textbf{73.83 $\pm$ 1.55}} & 13.16   \\
                    & & \multicolumn{1}{l|}{MSP} & EmoDB            & \multicolumn{1}{l|}{46.50 $\pm$ 1.08} & \multicolumn{1}{l|}{\textbf{66.17 $\pm$ 0.24}} & 19.67   \\ \hline \hline
\multirow{4}{*}{w/ noise} &\multirow{2}{*}{CC} & \multicolumn{1}{l|}{IEM} & ESD + noise    & \multicolumn{1}{l|}{62.30 $\pm$ 1.35} & \multicolumn{1}{l|}{\textbf{63.62 $\pm$ 1.65}} & 1.32   \\
                    & & \multicolumn{1}{l|}{MSP} & ESD + noise    & \multicolumn{1}{l|}{53.84 $\pm$ 1.41} & \multicolumn{1}{l|}{\textbf{57.77 $\pm$ 1.46}} & 3.93   \\  \cline{2-7}
& \multirow{2}{*}{CL} & \multicolumn{1}{l|}{IEM} & EmoDB + noise  & \multicolumn{1}{l|}{56.79 $\pm$ 3.09} & \multicolumn{1}{l|}{\textbf{68.50 $\pm$ 3.63}} & 11.71   \\
                    & & \multicolumn{1}{l|}{MSP} & EmoDB + noise  & \multicolumn{1}{l|}{60.22 $\pm$ 3.26} & \multicolumn{1}{l|}{\textbf{63.67 $\pm$ 1.65}} & 3.45   \\ \hline
\end{tabular}
\renewcommand{\arraystretch}{1.0}
\end{table*}

\subsection{Experiments simulating live data feed and real-world data scenarios}
\label{sec:eval_exp_3}

This study aims to compare the performance of supervised learning and reinforcement learning approaches in a simulated live data feed scenario. To avoid the difficulty of testing in a real-world environment, we continually push data from the target dataset and simulate a live scenario. The RL-DA approach continuously updates the model based on the incoming data, while the SL-DA method uses a static model trained on the source dataset.

The methodology of RL optimisation is similar to the experiment described in Section~\ref{sec:eval_exp_1}. First, the baseline model (SL-DA - rw) is trained with a subset of the source dataset. Then, the testing accuracy is measured using testing subsets of the target dataset.

We also evaluate the performance of RL-DA on real-world audio data. To achieve this, we mix background audio with the target datasets to create a new dataset with background noise. The DEMAND dataset's kitchen environment audio is used as the background sound to mix with the ESD and EmoDB datasets, with a Signal-to-Noise ratio of -5 dB.

Table~\ref{tab:baselineVsRealworld} shows the testing accuracy of the models trained using three different approaches (SL-DA - rw, RL-DA - target dataset without background noise (RL-DA w/o noise), and RL-DA - target dataset with background noise (RL-DA w/ noise)). The increment of accuracy in the adjacent RL approach compared to the SL approach is represented in the $\Delta$ columns.

Our findings indicate that RL-DA outperforms the baseline SL strategy in the live data feed scenario. In supervised learning methodologies, without a labelled dataset to retrain during inference time, the model cannot adapt to the deviation of a domain distribution. On the other hand, RL-based learning methodologies constantly receive feedback from the environment and optimise the agent model for domain deviations on the go. Retraining the model using supervised learning is feasible, but this would necessitate manual intervention. In contrast, RL-based methods incorporate this capability by design. Thus, RL is more suitable for domain adaptation in real-world scenarios due to its adaptability during optimisation.

\begin{figure}[th]
\centering
\includegraphics[width=0.45\textwidth]{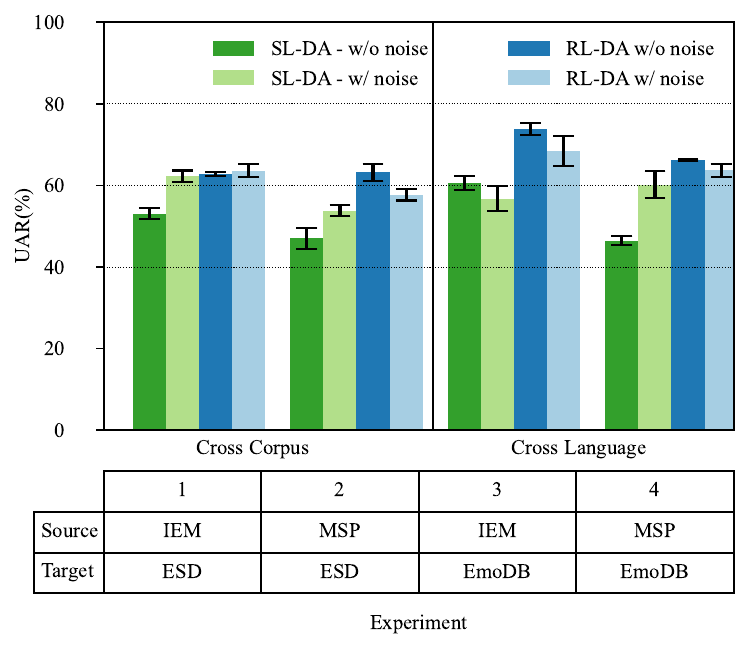}
\caption{Comparison of UAR of SL-DA - rw and the proposed RL-DA approach, without (w/o) background noise and with (w/) background noise of each cross-corpus and cross-language schema simulating live data feed scenario.}
\label{fig:baselineVsRealworld}
\end{figure}

Figure~\ref{fig:baselineVsRealworld} presents a comparison of the UAR achieved by the Baseline, RL-DA without noise, and RL-DA with noise in the simulated live data feed experiments. The results indicate that both RL-DA approaches outperform the baseline method.

To further evaluate the performance of RL-DA, we calculate the average improvement gained by using RL compared to the SL approach ($\mu_\Delta$) for each Cross-Corpus (${\mu_\Delta}^{CC}$) and Cross-Language (${\mu_\Delta}^{CL}$) schema using the equation~\ref{eq:average_increment}:
\begin{equation}
{\mu_\Delta}^{X} = \frac{\sum{\Delta^X}}{n} ;\ where\ X={CC\ or\ CL}
\label{eq:average_increment}
\end{equation}
Here, $n=4$, as there are 4 experiments for each CC and CL schema. The results show that using RL over the supervised learning approach yields an average increment ($\mu_\Delta$) of 11.77\% and 14.46\% for CC and CL schemas, respectively.

\subsection{Experiments with Wav2Vec2 input features}
\HL{
We conducted an additional experiment using the IEMOCAP dataset as the source and the ESD dataset as the target, employing wav2vec 2.0}~\cite{Baevski2020Wav2vecRepresentations}\HL{ as the feature extractor for raw audio instead of the MFCC input. The results are tabulated on Table}~\ref{tab:wav2vec_results}\HL{. We compare the results of the corresponding experiments with MFCC as feature. 
The average improvement of RL-DA over SL-DA ($\mu_\Delta$) was $10.44$, which aligns with the performance gains observed for the same source/target combination when using MFCC as the feature set as mentioned in Table}~\ref{tab:baselineVsRealworld}.
\begin{table}[h]
\caption{Comparison of UAR\% and UAR\% increment over SL-DA vs RL-DA approaches for the MFCC and Wav2Vec features. }
\label{tab:wav2vec_results}
\centering
\renewcommand{\arraystretch}{1.2}
\setlength\tabcolsep{4.5pt} 
\begin{tabular}{l|ll|ccl}
\hline
\multicolumn{1}{c|}{\multirow{2}{*}{\textbf{Feature}}} & \multicolumn{2}{c|}{\textbf{Dataset}}                  & \multicolumn{3}{c}{\textbf{UAR(\%)}}                                                    \\
\multicolumn{1}{c|}{}                                  & \multicolumn{1}{c|}{\textbf{Source}} & \multicolumn{1}{c|}{\textbf{Target}} & \multicolumn{1}{c|}{\textbf{SL-DA}} & \multicolumn{1}{c|}{\textbf{RL-DA}} & \multicolumn{1}{c}{$\mu_\Delta$} \\ \hline
MFCC                                                   & \multicolumn{1}{l|}{IEM}             & ESD             & \multicolumn{1}{c|}{53.11 $\pm$ 1.45}          & \multicolumn{1}{c|}{62.89 $\pm$ 0.51}          & 9.78        \\
Wav2Vec                                               & \multicolumn{1}{l|}{IEM}             & ESD             & \multicolumn{1}{c|}{57.63 $\pm$ 0.59}          & \multicolumn{1}{c|}{68.08 $\pm$ 1.65}          & 10.44  \\ \hline    
\end{tabular}
\renewcommand{\arraystretch}{1.0}
\end{table}

\HL{While wav2vec 2.0 increases computational resource demands and extends processing time, the primary focus of this study is on the relative performance difference between RL-DA and SL-DA, rather than on absolute accuracy. For researchers primarily interested in optimising accuracy, feature extractors like wav2vec 2.0 or similar alternatives may be preferable over MFCC.}

\subsection{Speaker Independent Experiments}
\label{sec:eval_exp_4}
This study extends our investigation by evaluating the proposed method for speaker-independent speech emotion recognition. We conduct two separate experiments using IEMOCAP and MSP-Improv as source datasets and ESD as the target dataset. To achieve speaker independence, we pre-train the base model using the source dataset and optimise it using the target dataset for the new domain. We evaluate the speaker-independent performance of the model by using a subset of the target dataset recorded by a specific speaker as the testing data and the remaining data as the training data.

\begin{figure}[ht]
\centering
\includegraphics[width=0.48\textwidth]{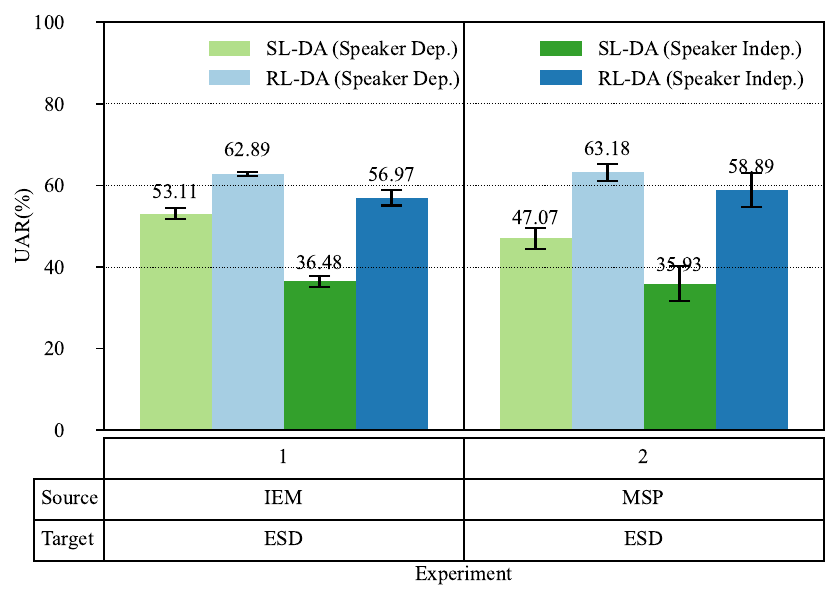}
\caption{Comparison of UAR of Speaker Dependent SL-DA, Speaker Dependent RL-DA, \HL{Speaker Independent SL-DA,} and Speaker Independent RL-DA.}
\label{fig:speaker_independent}
\end{figure}

We optimise the model with the training data and measure the testing accuracy. The experimental results are presented in Figure~\ref{fig:speaker_independent}. Comparing the results to Table~\ref{tab:baselineVsRealworld}, we observe that the RL-DA approach outperforms the baseline testing accuracy in both speaker-dependent and speaker-independent scenarios.

\HL{Additionally, it is apparent that in both SL-DA and RL-DA scenarios, speaker dependent performance is higher than speaker independent performance, a finding that is further supported by the literature.}~\cite{Rybka2013ComparisonRecognition}. 



%% file: 45_discussion.tex
\section{Discussion}
\label{sec:discussion}

\subsection{The Limitations of Traditional Supervised Learning in Real Time Adaptation}
Traditional SL models typically assume that the underlying data distribution remains static after training. However, in dynamic, real-world scenarios, the input data distribution changes due to various factors like environmental noise, different speakers, or new languages. Once an SL model is trained, it no longer adapts to these evolving conditions unless manually retrained. This rigidity is a major limitation when applying SL to tasks like SER in real-world settings.
In contrast, RL offers a solution by allowing models to learn from interactions with the environment continuously. While SL performs predictions based on a fixed mapping learned during training, RL is inherently dynamic, updating its policy as it receives new feedback, thereby improving its decision-making over time. This capability allows RL-based models to adapt to changing input distributions in real time, making them more suitable for real-world applications like SER.

The fundamental difference between SL and RL lies in their approach to learning. SL operates in a single-step process, where the model predicts outputs based on the training data it has seen, without adjusting to new data after deployment. In contrast, RL requires multiple iterative steps, where the agent interacts with the environment, receives feedback, and updates its policy continuously. While this makes RL more adaptive, it also leads to increased computational complexity and training time.

\subsection{Impact of Feedback Frequency on Model Adaptability}
\HL{In this study, we found that the frequency of feedback plays a critical role in the model’s ability to adapt effectively to new domains. Optimal performance is achieved when the model receives feedback for each prediction, as this continuous reinforcement allows for accurate adaptation to the shifting contexts encountered in real-world applications. However, we recognise that this setup may not always be feasible due to practical constraints. As such, periodic feedback can also be employed, although this may slow the rate of convergence and reduce the model’s adaptability over time. Ultimately, the decision on feedback frequency can be adjusted according to implementation needs, allowing for a balance between computational efficiency and model performance. Our findings reinforce the adaptability of reinforcement learning-based domain adaptation, highlighting how feedback frequency influences convergence rates and the overall adaptability of speech emotion recognition models in dynamic settings.}

\subsection{Limitations of Reinforcement Learning}
Despite the advantages of RL in handling dynamic environments, it has its own limitations. One major concern is stability. While RL is generally more robust to changing input distributions than SL, ensuring stability during training can be challenging. RL algorithms, especially deep RL methods, are known for their sensitivity to hyperparameters and exploration strategies. Poor exploration can lead to sub optimal policies, and over-exploration may result in instability during the training phase~\cite{Mnih2015Human-levelLearning}. 

\HL{An additional limitation in this approach is the observed variability in performance outcomes, as indicated by a higher standard deviation in the results. This higher variability suggests that the model may lack stability, which can hinder its reliability in certain applications. Such fluctuations in performance underscore the sensitivity of RL to specific conditions and further highlight the potential instability introduced during training. This issue needs to be addressed to ensure consistent model behaviour and to reduce the risk of unreliable performance in real-world applications.}

Additionally, while RL's adaptability is beneficial, it often requires many iterations to converge, increasing computational costs and training time~\cite{Li2017DeepOverview}. This presents a trade-off between the model's ability to adapt and the resources required to maintain such flexibility. Although RL shows promise for domain adaptation, particularly in real-world SER tasks, these limitations must be carefully managed to fully realise its potential.

%% file: 60_conclusion.tex
\section{Conclusion}


This study investigated the use of Reinforcement Learning (RL) for domain adaptation in Speech Emotion Recognition (SER), marking the first attempt to apply RL for this purpose in SER. Previous approaches have primarily relied on supervised learning (SL) techniques, but the findings of this study suggest that RL offers superior domain adaptability, particularly in real-world scenarios. The proposed RL-based domain adaptation (RL-DA) framework demonstrated an increase in accuracy by 11\% and 14\% in cross-corpus and cross-language setups, respectively, compared to baseline SL approaches.

A key factor contributing to this improved performance is the RL agent's ability to continually incorporate feedback, eliminating the need for manual retraining that SL models require. Furthermore, the study introduced a simulated environment for training, given the challenges of implementing a human-in-the-middle system during the training phase. Additionally, a human-in-the-middle application has been developed and publicly hosted at~\url{https://rlemotion.cloud.edu.au}, showcasing practical implementation in dynamic real-world scenarios.

Future directions will focus on enhancing the framework through continuous learning and scenario-specific testing. By enabling online training, the model can dynamically adjust to new conditions without full retraining. Scenario-specific protocols involving real-time interactions in varied settings will also be explored, enabling robust performance across different environments such as noisy versus quiet spaces or indoor versus outdoor contexts.

The outcomes of this research set a new direction for the community and pave the way for developing emotion-aware applications that can flexibly adapt to users' emotional expressions across diverse domains.

%% file: 50_biography.tex
\begin{IEEEbiography}
	[{\includegraphics[width=1in,height=1.25in,clip,keepaspectratio]{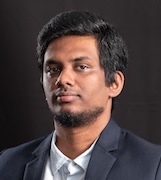}}]
	{Thejan Rajapakshe} (Student Member, IEEE) received a bachelor's degree in Applied Sciences from the Rajarata University of Sri Lanka and a Bachelor of Information Technology from the University of Colombo, School of Computing, in 2016.
 
 He is currently a research scholar at the University of Southern Queensland (UniSQ). His research interests include reinforcement learning, speech processing, and deep learning.
\end{IEEEbiography}

\begin{IEEEbiography}
	[{\includegraphics[width=1in,height=1.25in,clip,keepaspectratio]{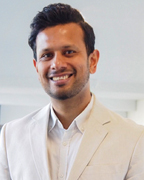}}]
	{Rajib Rana} (Member, IEEE) received his Ph.\,D.\ in Computer Science and Engineering from the University of New South Wales, Sydney, Australia, in 2011.
 
	He is an experimental computer scientist, and a Professor of Computer Science in the University of Southern Queensland. He is also the Director of the IoT Health research program at the University of Southern Queensland. He is a recipient of the prestigious Young Tall Poppy QLD Award 2018 as one of Queensland’s most outstanding scientists for achievements in the area of scientific research and communication. Rana's research work aims to capitalise on advancements in technology along with sophisticated information and data processing to better understand disease progression in chronic health conditions and develop predictive algorithms for chronic diseases, such as mental illness and cancer. His current research focus is on Unsupervised Representation Learning, Reinforcement Learning, Adversarial Machine Learning and Emotional Speech Generation. He received his B.\,Sc.\ degree in Computer Science and Engineering from Khulna University, Bangladesh, with the Prime Minister and President's Gold Medal for outstanding achievements. He received his postdoctoral training at Autonomous Systems Laboratory, CSIRO, before joining the University of Southern Queensland  in 2015.
\end{IEEEbiography}

\begin{IEEEbiography}
	[{\includegraphics[width=1in,height=1.25in,clip,keepaspectratio]{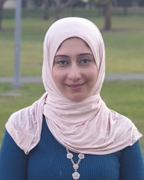}}]
	{Sara Khalifa} (Member, IEEE) received her Ph.\,D.\ in Computer Science and Engineering from UNSW (Sydney, Australia).
 
	She is currently an associate professor at Queensland University of Technology. 
 She is also an honorary adjunct lecturer at the University of Queensland and conjoint lecturer at the University of New South Wales.  Her research interests rotate around the broad aspects of mobile and ubiquitous computing, mobile sensing, and Internet of Things (IoT). Her PhD dissertation received the 2017 John Makepeace Bennett Award which is awarded by CORE (Computing Research and Education Association of Australasia) to the best PhD dissertation of the year within Australia and New Zealand in the field of Computer Science. Her research has been recognised by multiple awards including 2017 NSW Mobility Innovation of the year, 2017 NSW R\&D Innovation of the year, National Merit R\&D Innovation of the year, and the Merit R\&D award at the Asia Pacific ICT Alliance (APICTA) Awards, commonly known as the `Oscar' of the ICT industry in the Asia Pacific, among others.
\end{IEEEbiography}

\begin{IEEEbiography}
	[{\includegraphics[width=1in,height=1.25in,clip,keepaspectratio]{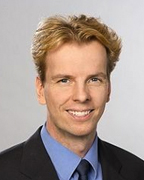}}]
	{Bj\"{o}rn W.\ Schuller} (Fellow, IEEE) received his Diploma in 1999, his doctoral degree for his study on Automatic Speech and Emotion Recognition in 2006, and his habilitation and Adjunct Teaching Professorship in the subject area of Signal Processing and Machine Intelligence in 2012, all in electrical engineering and information technology from TUM in Munich/Germany.
 
 He is a Professor of Artificial Intelligence in the Department of Computing at the Imperial College London/UK, where he heads GLAM — the Group on Language, Audio, \& Music, Full Professor and head of the Chair of Embedded Intelligence for Health Care and Wellbeing at the University of Augsburg/Germany, and CEO/CSO of audEERING. He was previously full professor and head of the Chair of Complex and Intelligent Systems at the University of Passau/Germany. Professor Schuller is Fellow of the IEEE, Golden Core Member of the IEEE Computer Society, Fellow of the ISCA, Fellow of the BCS, Senior Member of the ACM, and Fellow and President-emeritus of the Association for the Advancement of Affective Computing (AAAC). He (co-)authored 5 books and 1200+ publications in peer-reviewed books, journals, and conference proceedings leading to more than overall 48\,000 citations (h-index = 99). Schuller was general chair of ACII 2019, Program Chair of Interspeech 2019 and ICMI 2019, repeated Area Chair of ICASSP, and former Editor in Chief of the IEEE Transactions on Affective Computing, and is Field Chief Editor of the Frontiers in Digital Health.
\end{IEEEbiography}